\documentclass[nofootinbib,aps,11pt,preprintnumbers,unsortedaddress,prd]{revtex4}
\usepackage{graphicx}
\usepackage{cancel}
\usepackage{amssymb}
\usepackage{textcomp}
\usepackage{amsmath}
\usepackage{bm}
\usepackage{times}
\usepackage{epsfig}
\usepackage{color}
\usepackage{graphics}
\usepackage{hyperref}
\usepackage{setspace}
\usepackage{epsfig}
\usepackage{amsmath}
\usepackage{bm}
\usepackage{times}
\usepackage{graphicx}
\usepackage{color}
\usepackage{slashed}
\usepackage{graphicx}
\usepackage{amsmath, amssymb}

\hypersetup{
    pdfnewwindow=true,      
    colorlinks=true,       
    linkcolor=black,          
    citecolor=blue,        
    filecolor=blue,      
    urlcolor=blue           
}

\def\bea{\begin{eqnarray}}
\def\eea{\end{eqnarray}}
\begin{document}
\title{\Large {\it{\bf{Minimal Theory for Lepto-Baryons}}}}
\author{Pavel Fileviez P\'erez}
\author{Sebastian Ohmer}
\author{Hiren H. Patel}
\affiliation{\vspace{0.15cm} \\  Particle and Astro-Particle Physics Division \\
Max-Planck Institute for Nuclear Physics {\rm{(MPIK)}} \\
Saupfercheckweg 1, 69117 Heidelberg, Germany}
%
\date{\today}
\begin{abstract}
We investigate the simplest models where baryon and lepton numbers are defined as local symmetries spontaneously broken at the low scale 
and discuss the implications for cosmology. We define the simplest anomaly-free theory 
for spontaneous baryon and lepton number symmetry breaking which predicts the existence of lepto-baryons. In this context,  we study the new sphaleron condition on the chemical potentials 
and show the relation between the present baryon asymmetry and the $B-L$ asymmetry generated in the early universe. 
The properties of the cold dark matter candidate for which stability is a natural consequence from symmetry breaking are briefly discussed. 
\end{abstract}
\maketitle

\section{Introduction}
\label{intro}
One of the bonuses of the Standard Model is the classical conservation of baryon and lepton numbers, which forbids proton decay. Unfortunately, when high-dimension operators~\cite{Weinberg} consistent with gauge symmetries of the Standard Model are considered, these global symmetries are lost.  Then, one needs to postulate a suppression mechanism 
to be consistent with the experimental bounds on the proton lifetime~\cite{nath}.
An example is the existence of the great desert with a new scale related to the grand unification scale of approximately $M_{GUT} \sim 10^{14-16}$ GeV~\cite{Georgi}.

Recently, the authors of Ref.~\cite{Wise1} proposed an alternative approach for physics beyond 
the Standard Model where the baryonic and leptonic symmetries are considered fundamental 
local symmetries. In this context, these symmetries can be spontaneously broken near the electroweak scale and there is hope to test the predictions at colliders. The original idea related to the spontaneous breaking of local baryon number with the use of the Higgs mechanism was proposed by Pais ~\cite{Pais}.  In the models investigated in 
Refs.~\cite{Wise1,Wise2,Wise3} local baryon number is always broken in three units and the 
proton is stable. There are also interesting cosmological implications since the new sector typically contains a cold dark matter candidate 
for which stability is automatic, and also the usual picture for baryogenesis can be different~\cite{Duerr, Patel}.

In this article, we revisit the idea of constructing a simple particle physics model where baryon and lepton numbers are local symmetries.
We find that the simplest phenomenologically viable model that is free of anomalies consists of three right-handed neutrinos and only four new fermion multiplets, which we call ``lepto-baryons''. We investigate the main features of this theory and discuss the main implications for cosmology, including the modified sphaleron conditions on the chemical potentials.  After symmetry breaking 
the new fermions have the same quantum numbers as the gauginos and higgsinos in the Minimal Supersymmetric Standard Model. 
In this model the dark matter candidate is the lightest Majorana fermion with baryon number.

\section{Theories of Local Baryon and Lepton Numbers}
In this section, we systematically show that the simplest viable model with gauged baryon and lepton numbers requires four new fermion multiplets along with three right-handed neutrinos to cancel all anomalies.  Here, we follow the discussion in Ref.~\cite{Wise1} to understand the anomaly cancellation.  Our main goal is to construct a theory based on the gauge group
$$G_{BL}=SU(3)_C \otimes SU(2)_L \otimes U(1)_Y \otimes U(1)_B \otimes U(1)_{\ell}.$$ 
To simplify the discussion we first search for minimal models with gauged baryon number, and later show how the methods easily extend to the case where lepton number is also a local symmetry.

\subsection{Gauging baryon number}
\label{theories baryon}
In order to consistently gauge baryon number, one has to introduce a new set of fermions to cancel the non-trivial baryonic anomalies 
$$\mathcal{A}^{SM}_1 (SU(2)^2_L \otimes U(1)_B) = 3/2  \enspace \text{and} \enspace  \mathcal{A}^{SM}_2 (U(1)_Y^2 \otimes U(1)_B) = -3/2.$$
In addition, the vanishing of the other Standard Model baryonic anomalies
\begin{align*}
\mathcal{A}^{SM}_3 (SU(3)_C^2 \otimes U(1)_B)=0, &\enspace \mathcal{A}^{SM}_4 (U(1)_Y \otimes U(1)^2_B) = 0, \\
\mathcal{A}_5^{SM} (U(1)_B) = 0, &\enspace  \mathcal{A}_6^{SM} (U(1)^3_B) = 0,
\end{align*}
along with the gauge anomalies must be maintained upon the introduction of additional fermions.  The anomaly $\mathcal{A}^{SM}_3=0$ implies that the simplest solution for anomaly cancellation corresponds to cases with colorless fields.  Therefore, we proceed with fields without color.

We begin our discussion with the introduction of just two new fermionic multiplets
\begin{equation*}
\Psi_L \sim (1, N, y_1, B_1) \enspace  \text{and}  \enspace \Psi_R \sim (1, M, y_2, B_2).
\end{equation*}
The anomaly $\mathcal{A}_5(U(1)_B) = 0$ requires $B_1 = M B_2 / N$. Then, using $\mathcal{A}_6 (U(1)^3_B) = 0$, the condition $M = N$ holds and hence $B_1 = B_2$. 
Then,  $\mathcal{A}_1 (SU(2)^2_L \otimes U(1)_B)$ cannot be canceled. Thus, there is no solution with just two new fermionic fields.

Next, we look for the solution with three multiplets 
\begin{align*}
\Phi_L \sim (1 , N, y , B_1 ), \quad \Phi_R \sim (1, N, y, B_2 )\quad \text{and} \quad \chi_L \sim (1, M, 0, B_3)  \,.
\end{align*}
Cancellation of anomalies requires $M = 2N$, $y^2 = N^2/4$ and $B_1 = -B_2 = -B_3= 3 / N^3$.  That $\chi_L$ is an even dimensional isospin multiplet with zero hyperchange implies the existence of fermions with fractional electric charge, and thus cannot decay into Standard Model particles. More general quantum number assignments
will lead to the same conclusion. Therefore, models with three fermion multiplets, while consistent with respect to anomaly cancellation, requires a new stable electrically charged particle which is ruled out by cosmology.  

Finally, we turn to the case of four fermion multiplets, with the additional fermion field denoted as $\Sigma_L$.  We find that that the simplest solution that satisfies anomaly cancellation is
\begin{align*}
\Psi_L &\sim (1,\, 2,\, 1/2,\, 3/2),  &\Psi_R &\sim (1,\, 2,\, 1/2,\, -3/2), \nonumber \\ 
\Sigma_L &\sim (1,\, 3,\, 0,\, -3/2), &   \chi_L &\sim (1,\, 1,\, 0,\, -3/2) \,.
\end{align*}
The charge assignments of the new fermions does not lead to fractional charges in the particle spectrum, so that the model is not immediately ruled out by cosmology.  To summarize, we observe that the requirement of anomaly cancellations always leads to particle content which allows for vector-like masses for the new fields, and that the simplest viable solutions tend to correspond to models with an even number of colorless multiplets.  There exists a non-minimal solution where the fields transform under the adjoint representation of the $SU(3)_C$.
%
\subsection{Minimal theory of lepto-baryons}
%
Knowing a solution for gauging baryon number, we follow the prescription described in Ref.~\cite{Wise1} to construct the model in which lepton number is also a local symmetry.  
Adding the three right-handed neutrinos the linear and cubic leptonic anomalies are cancelled.  As noticed in Ref.~\cite{Wise1}, there is a symmetry between the anomaly cancellation 
in the quark and lepton sectors.  Now, the anomalies to be cancelled are
$$\mathcal{A}^{SM}_7 (SU(2)^2_L \otimes U(1)_{\ell}) = 3/2 \  \text{and} \ \mathcal{A}^{SM}_8 (U(1)_Y^2 \otimes U(1)_{\ell}) = -3/2.$$
The other anomalies are automatically zero in the Standard Model with right-handed neutrinos, 
i.e. $$\mathcal{A}^{SM}_9 (SU(3)_C^2 \otimes U(1)_{\ell})=0, \mathcal{A}^{SM}_{10} (U(1)_Y \otimes U(1)^2_{\ell}) = 0, 
\mathcal{A}_{11}^{SM} (U(1)_{\ell}) = 0, \  \text{and} \  \mathcal{A}_{12}^{SM} (U(1)^3_{\ell}) = 0,$$ and
in general one also must cancel the mixed anomalies
$$\mathcal{A}_{13} (U(1)_B^2 \otimes U(1)_{\ell})=0, \mathcal{A}_{14} (U(1)_B \otimes U(1)^2_{\ell}) = 0, \  \text{and} \
\mathcal{A}_{15} (U(1)_Y \otimes U(1)_B \otimes U(1)_{\ell} ) = 0.$$
Once new fields are included in the theory, the cancellation must be maintained. The simplest solution is to assign to the new fields the same lepton charges as their baryon charges. Therefore, the new fields are
\begin{align}
\Psi_L \sim (1 , 2 , 1/2 , 3/2, 3/2 ), \quad \ \Psi_R \sim (1, 2, 1/2, -3/2,-3/2 ), \nonumber \\ 
\Sigma_L \sim (1, 3, 0, -3/2,-3/2),  \quad \ \text{and} \  \quad \chi_L \sim (1, 1, 0, -3/2,-3/2) \,.
\end{align}
We call these fields ``lepto-baryons" and in component form they can be expressed as
\begin{align}
\Psi_{L} = \begin{pmatrix} \psi^+_{1} \\ \psi^0_{1} \end{pmatrix}, 
\quad  
(\Psi^c)_{L} = (\Psi_R)^c=\begin{pmatrix} \psi^0_{2}  \\  \psi^-_{2} \end{pmatrix}, 
\quad 
\text{and} \quad \Sigma_L = \frac{1}{2} \begin{pmatrix} \Sigma^0 && \sqrt{2} \Sigma^+ \\ \sqrt{2}\Sigma^- && - \Sigma^0 \end{pmatrix} \,.
\end{align}
The relevant interactions in this model are given by
\begin{eqnarray}
-\mathcal{L}  & \supset  & h_1  \bar{\Psi}_R  H \chi_L + h_2  H^\dagger \Psi_L \chi_L 
+  h_3  H^\dagger \Sigma_L  \Psi_L  + h_4  \bar{\Psi}_R  \Sigma_L H \nonumber \\
&& \hspace{2mm}+ \, \lambda_\Psi \ \bar{\Psi}_R \Psi_L S_B^* + \lambda_\chi \ \chi_L \chi_L S_B + \lambda_{\Sigma} \ \text{Tr} \ \Sigma_L^2 S_B \nonumber \\
& &  \hspace{2mm} + \,Y_\nu \ \ell_L H \nu^c + \lambda_R  \  \nu^c \nu^c S_L \ + \   \text{h.c.},
\end{eqnarray}
where $\nu^c=(\nu_R)^c$ are the right-handed neutrinos. The needed scalar sector is composed of the fields
\begin{align}
S_B \sim (1, 1, 0, 3, 3), \  S_L \sim (1, 1, 0, 0, 2) \ \text{and} \ H \sim (1, 2, 1/2, 0, 0),
\end{align}
which are responsible for spontaneous symmetry breakdown of the baryonic, leptonic, and electroweak symmetries, respectively.
We define the vacuum expectation values as $\langle H \rangle = v / \sqrt{2}$, $\langle S_B \rangle = v_B / \sqrt{2}$ and  $\langle S_L \rangle = v_L / \sqrt{2}$.
In order to understand the implications for baryogenesis we will assume that $U(1)_{\ell}$ is broken at a scale much larger than the electroweak-scale. When $S_B$ acquires a vacuum expectation value, the local baryonic symmetry $U(1)_B$ is broken to the $Z_2$ symmetry which guarantees the stability of the 
lightest lepto-baryon. This will be our dark matter candidate. 

After symmetry breaking the mass matrix for the neutral fermions in the basis $(\psi^0_1, \psi^0_2, \Sigma^0 , \chi^0)$ is given by
\begin{align}
{\cal M}_0=\begin{pmatrix}
0 & \frac{1}{\sqrt{2}} \lambda_\Psi v_B & -\frac{1}{2 \sqrt{2}} h_3 v & \frac{1}{\sqrt{2}}h_2 v  \\
\frac{1}{\sqrt{2}} \lambda_\Psi v_B  & 0 & - \frac{1}{2 \sqrt{2}} h_4 v & \frac{1}{\sqrt{2}} h_1 v \\
-\frac{1}{2 \sqrt{2}} h_3 & - \frac{1}{2 \sqrt{2}} h_4 v & \frac{1}{\sqrt{2}} \lambda_\Sigma v_B& 0\\
\frac{1}{\sqrt{2}}h_2 v  &  \frac{1}{\sqrt{2}} h_1 v & 0 & \sqrt{2} \lambda_\chi v_B \\
\end{pmatrix} ,
\end{align}
and the mass matrix for the charged fermions in the basis $(\psi_1^+, \Sigma^+)$ and $(\psi_2^-, \Sigma^-)$ reads
\begin{align}
{\cal M}_\pm=\begin{pmatrix}
- \frac{1}{\sqrt{2}} \lambda_\Psi v_B & \frac{1}{2} h_3 v  \\
- \frac{1}{2} h_4 v &  \frac{1}{\sqrt{2}} \lambda_\Sigma v_B \\
\end{pmatrix}\,.
\end{align}
Notice that when all Yukawa couplings $h_i$ are small the new fermions have pure vector-like masses proportional to the vacuum expectation value of $S_B$. In this case, the model can easily satisfy constraints coming from electroweak precision observables, collider searches and others. Notice that these fields have the same quantum numbers as the gaugino-Higgsino fields in the Minimal Supersymmetric Standard Model.

As it has been pointed out in Ref.~\cite{Wise1}, this type of model is interesting because the symmetry breaking scale can be low and there is no need to postulate 
the existence of the great desert between the electroweak and Planck scales. The main reason is that the scalar field $S_B$ responsible for breaking baryon number carries baryonic charge +3, and so does not generate any contribution to proton decay.  There are higher-dimension operators that would mediate baryon number-violating decays by three units.  For example, one could consider
\begin{align}
\frac{c}{\Lambda^{15}} \left(  QQQL \right)^3 S_B^*,
\end{align}
which mediates the processes: $ p p p \to e^+ e^+ e^+$, $p p n \to e^+ e^+  \bar{\nu}$, $p n n \to  e^+ \bar{\nu} \bar{\nu}$, and
$3 n \to 3 \bar{\nu}$. However, even when $\Lambda$ is of order 1 TeV, these processes are highly suppressed due to the large power $\Lambda^{15}$.  It is important to mention that the vacuum expectation value, $v_B$, sets the scale for the mass of the leptophobic gauge 
boson $Z_B$, and bounds on the mass of this type of gauge bosons are weak~\cite{Dobrescu,An,ATLAS}.

\section{Baryon Asymmetry, Sphalerons and Dark Matter}
\label{baryon and sphaleron}
In this section, we investigate the relation between the baryon asymmetry and the initial $B-L$ asymmetry by solving the chemical equilibrium equations.  Below, we will briefly discuss how the $B-L$ asymmetry is generated in this model. Here, we will assume that the local leptonic symmetry is broken far above the electroweak scale.
The baryon asymmetry is given by
\begin{equation}
B_f = \frac{n_q - n_{\bar{q}}}{s} = \frac{15}{4 \pi^2 g_* T} \times 3 \left( \mu_{u_L} + \mu_{u_R} + \mu_{d_L} + \mu_{d_R} \right),
\end{equation}
when the chemical potentials are much smaller than the temperature, $\mu \ll T$. The quantity $g_*$ is the total number of relativistic degrees of freedom and $s$ is the entropy density. 

The only conserved global symmetry after symmetry breaking in the theory is the usual $B-L$ in the Standard Model 
and we will use it to determine the final baryon asymmetry. The $B-L$ asymmetry is defined by
\begin{align}
\Delta(B-L)_\text{SM} = \frac{15}{4\pi^2 g_* T} \times 3 (\mu_{u_L} + \mu_{u_R} +\mu_{d_L} +\mu_{d_R} -\mu_{\nu_L} -\mu_{e_L} -\mu_{e_R})\,.
\end{align}
Assuming isospin conservation one has the conditions~\cite{Harvey} on the chemical potentials
\begin{align}
\mu_{u_L} = \mu_{d_L}, \qquad \mu_{e_L} = \mu_{\nu_L}, \qquad  {\text{and}} \qquad \mu_0=\mu_+ \,.
\end{align}
Here $\mu_0$ and $\mu_+$ are the chemical potentials of the neutral and charged components of the SM Higgs doublet. The Standard Model interactions for quarks and leptons with the Higgs give us the useful relations
\begin{align}
\mu_{u_R} &= \mu_0 +\mu_{u_L}, \quad
\mu_{d_R} = - \mu_0 +\mu_{d_L}, \quad \text{and} \quad
\mu_{e_R} = -\mu_0 + \mu_{e_L} \,.
\end{align}
The interactions proportional to $\lambda_i$ give the following new relations
\begin{align}
2\mu_{\chi_L} + \mu_{S_B} &= 0, \quad
2 \mu_{\Sigma_L} + \mu_{S_B} = 0 \quad \text{and} \quad
-\mu_{\Psi_L} + \mu_{\Psi_R} + \mu_{S_B} = 0, 
\end{align}
while the $h_i$ interactions impose the conditions  
\begin{align}
-\mu_{\Psi_R} + \mu_{0} + \mu_{\chi_L} &= 0, \quad
-\mu_{0} + \mu_{\Psi_L } + \mu_{\chi_L} = 0, \nonumber \\
- \mu_{0} + \mu_{\Sigma_L} + \mu_{\Psi_L} &= 0, \quad {\text{and}} \
-\mu_{\Psi_R} + \mu_{\Sigma_L} + \mu_0 = 0 \,.
\end{align}
The sphaleron condition must conserve the total baryon number.  Because the lepto-baryons carry $SU(2)_L$ quantum numbers, they are expected to contribute to the effective 't Hooft operator.  In Ref.~\cite{Creutz:2007yr}, the 't Hooft vertex for arbitrary representations was derived iteratively.  
Here, we present a pedestrian approach to arrive at the same result. The basic idea is to integrate the anomaly 
equation to determine the change in the number of quanta corresponding to an instanton of unit winding number.  
Consider a left-handed fermion transforming as an isospin $j$ representation under $SU(2)_L$.  The anomaly equation reads
\begin{equation}
\partial_\mu J^\mu =\frac{g^2}{16\pi^2}\mathcal{A}^{ab}W_{\mu\nu}^a \widetilde{W}^{\mu\nu\,b}\,,
\end{equation}
where $\mathcal{A}^{ab}=\text{Tr}\big[G {\textstyle\frac{1}{2}}\{T^a_j,\,T^b_j\}\big]$.
The current carries a normalizing abelian charge $G$. Upon inserting the $SU(2)$ identity $\text{Tr}\big(T_j^a T_j^b\big)=\frac{1}{3}j(j+1)(2j+1)\delta^{ab}$, we have
\begin{equation}
\partial_\mu J^\mu =\frac{g^2}{16\pi^2}G\frac{1}{3}j(j+1)(2j+1) \delta^{ab}W_{\mu\nu}^a \widetilde{W}^{\mu\nu\,b}.
\end{equation}
In order to integrate the continuity equation and convert the right hand side into a winding number,
we express the field strength tensors in terms of matrix-valued fields by writing the Kronecker delta as a trace over a product of $SU(2)$ generators in the fundamental representation, $\delta^{ab}=2\,\text{Tr}\big(T^a_{1/2} T^b_{1/2}\big)$, to obtain
\begin{equation}
\frac{\partial \rho_G}{\partial t}+\nabla\cdot\mathbf{J} = G\frac{2}{3}j(j+1)(2j+1)\left(\frac{g^2}{16\pi^2}\text{Tr}\big[W_{\mu\nu} \widetilde{W}^{\mu\nu}\big]\right)\,.
\end{equation}
Upon inserting the explicit form of the instanton into the right hand side, we may integrate both sides over the entire 4-volume. In the left hand side we get the change of $G$-charge, and on the right hand side the quantity inside the parenthesis yields the winding number, $\nu=1$, for the instanton configuration.  Upon performing the integration we arrive at
\begin{equation}
\frac{\Delta G}{G}=\Delta n_L = \frac{2}{3}j(j+1)(2j+1)\,,
\end{equation}
which is the change in the number of left-handed quanta corresponding to an instanton event, in agreement with Ref.~\cite{Creutz:2007yr}. 
The 't Hooft vertex for a given multiplet is schematically of the form: $\mathcal{L}_\text{eff}=(\psi_L)^{\Delta n_L}$.  Therefore, in this model the 't Hooft operator is
\begin{align}
(QQQL)^3 \bar{\Psi}_R \Psi_L \Sigma_L^4.
\end{align}
Notice that this operator is different from the one in the Standard Model because of the appearance of extra fields to conserve total baryon number.  Hence, the relevant equilibrium condition imposed by the sphaleron processes is
\begin{align}
3(3\mu_{u_L} + \mu_{e_L}) + \mu_{\Psi_L} - \mu_{\Psi_R} + 4 \mu_{\Sigma_L} = 0 \,.
\end{align}
Finally, using Eq.~(10) conservation of electric charge implies
\begin{align}
6 ( \mu_{u_L} + \mu_{u_R}) - 3 (\mu_{d_L} + \mu_{d_R}) - 3 (\mu_{e_L} + \mu_{e_R}) + 2 \mu_{0} + \mu_{\Psi_L} + \mu_{\Psi_R}= 0 \,,
\end{align}
while total baryon number conservation above the symmetry breaking scale gives us
\begin{align}
3 ( \mu_{u_L} + \mu_{u_R} + \mu_{d_L} + \mu_{d_R}) + \frac{3}{2} ( 2 \mu_{\Psi_L} - 2 \mu_{\Psi_R} - 3 \mu_{\Sigma_L} - \mu_{\chi_L})
+ 6 \mu_{S_B}= 0 \,.
\end{align}
Therefore, we arrive at the following system of equations
\begin{eqnarray}
\Delta(B-L)_\text{SM} &=& \frac{15}{4\pi^2 g_* T}(12 \mu_{u_L} - 9 \mu_{e_L} + 3\mu_0), 
\\
Q_{em} = 0 &\implies& 3\mu_{u_L} - 3 \mu_{e_L}  + 8 \mu_0 = 0, 
 \\
B_T = 0 &\implies&  \mu_{u_L} - 2 \mu_{\chi_L} = 0, 
\\
\text{Sphalerons} &\implies& 9 \mu_{u_L} + 3 \mu_{e_L} + 2 \mu_{\chi_L} = 0 \,.
\end{eqnarray}
With this, we can express the final baryon asymmetry as a function of the initial $B-L$ asymmetry 
\begin{align}
B_f = \frac{15}{4\pi^2 g_* T} (12 \mu_{u_L}) = \frac{32}{99} \Delta(B-L)_\text{SM} \approx 0.32 \ \Delta(B-L)_\text{SM} \,.
\end{align}
Notice that the conversion factor is different from the one in the Standard Model, i.e. smaller than $28/79\approx 0.35$~\cite{Harvey}.
As it is well-known this conversion factor is important to understand how the initial $B-L$ asymmetry is transferred into the baryon asymmetry 
without  assuming any particular model. Notice that in general there is no reason to expect a simple solution for the conversion factor in this model 
because the sphalerons conserve the total baryon number. In this way, we show that it is possible to have a consistent scenario for baryogenesis.

The primordial $B-L$ asymmetry can be generated by the out-of-equilibrium decays of the right-handed neutrinos as in canonical leptogenesis scenarios.  Once $U(1)_\ell$ is spontaneously broken a mass term for the right-handed neutrinos is generated.  The leptogenesis scale is required to be below the $U(1)_{\ell}$ breaking scale. We have assumed that this mechanism is realized at the high scale typical of leptogenesis scenarios (see \cite{Buchmuller:2004nz} for details).

In this model the lightest lepto-baryon is automatically stable. 
Here, we discuss briefly how to explain the observed relic density in the Universe in the case when the dark matter candidate 
is the $\chi$ Majorana field, with mass approximately equal to $\sqrt{2} \lambda_\chi v_B$. After symmetry breaking the three CP-even physical Higgses 
can have interactions with all Standard Model fields and the dark matter candidate. Therefore, we can have the following velocity suppressed annihilation channels,  
\begin{gather*} \chi \ \chi \  \to \ H_i \  \to \  \bar{q} q,\, h h,\, WW,\, ZZ,\,\ldots\,.
\end{gather*} 
through the Higgs portal.
From the interaction with the leptophobic gauge boson, the following additional channel
$$ \chi \ \chi \  \to \  Z_B \ \to \ \bar{q} q,$$ 
is open.  Unfortunately, this channel is velocity suppressed since $\chi$ is a Majorana fermion. However, the dark matter 
candidate can annihilate into two gauge bosons or into Higgs bosons through the $t$-channel
 $$ \chi \ \chi \  \to \  Z_B  Z_B, \ H_i H_j, \ {H_i Z_B},$$
which are not velocity suppressed. The direct detection of this dark matter candidate can be through the Higgs portal or the leptophobic gauge boson. 
The elastic nucleon-dark matter cross section mediated by the gauge boson is velocity suppressed and it is difficult to test the predictions at current experiments. 
The most optimistic scenario for placing constraints comes from the Higgs portal interaction. In this case, predictions for direct detection of a Majorana fermion have been studied by many authors in the field~\cite{DMreview}. The detailed analysis of this dark matter candidate will appear in a future publication.
\section{Summary}
\label{summary}
In this article, we have proposed the simplest extension of the Standard Model 
where baryon and lepton numbers can be defined as local symmetries spontaneously broken 
at the low scale. The new sector of the theory is composed of four new fermionic multiplets called ``lepto-baryons" and the right-handed neutrinos.  Baryon number is spontaneously broken in three units allowing for low scale symmetry breaking consistent with experimental bounds on proton lifetime.  Upon symmetry breaking the lepto-baryons have the same quantum numbers as the gaugino-Higgsino fields in the Minimal Supersymmetric Standard Model. Therefore, most of the phenomenological studies on gauginos and Higgsinos can be relevant for this model.

We have investigated the implications for cosmology when the local baryon number is broken near the electroweak scale.  In this case, the sphaleron 't Hooft operator is different, modifying the relationship between the initial $B-L$ asymmetry generated by a mechanism such as leptogenesis and the final baryon asymmetry. 
This model also predicts the existence of a cold dark matter candidate which is a Majorana fermion.  
The results presented in this article are complementary to the results in Refs.~\cite{Wise1,Wise2,Wise3}, 
because in this way we know which are the simplest extensions of the Standard Model with gauged baryon 
and lepton numbers. These theories provide a guide to find a unified theory at the low scale.
\\
\\
{\textit{Acknowledgments:}}
{\small {P. F. P.\ thanks M. B. Wise for many discussions and the theory group at Caltech for hospitality. The work of P. F. P.  is partly funded by the Gordon and Betty Moore Foundation through Grant 776 to the California Institute of Technology.}}



\end{document}